\documentclass[12pt, letterpaper]{article}
\usepackage[top=1in, bottom=1.5in, left=1in, right=1in]{geometry}
\usepackage[utf8]{inputenc}
\usepackage{booktabs}
\usepackage{hyperref}
\usepackage{graphicx}

\title{The Gaia-LSST Synergy: 
resolved stellar populations in selected Local Group stellar systems}
\author{\normalsize G. Clementini$^1$(gisella.clementini@inaf.it), I. Musella$^2$(ilaria.musella@inaf.it), \\\normalsize A. Chieffi$^3$(alessandro.chieffi@inaf.it), M. Cignoni$^4$(michele.cignoni@unipi.it), \\\normalsize F. Cusano$^1$(felice.cusano@inaf.it), M. Di Criscienzo$^5$(marcella.dicriscienzo@inaf.it), \\\normalsize M. Fabrizio$^{5,6}$(michele.fabrizio@inaf.it), A. Garofalo$^{1,7}$(alessia.garofalo@unibo.it), 
\\\normalsize S. Leccia$^2$ (silvio.leccia@inaf.it), M. Limongi$^5$(marco.limongi@inaf.it), \\\normalsize M.  Marconi$^2$(marcella.marconi@inaf.it), E. Marini$^{5,8}$(estermarini@hotmail.it ), 
\\\normalsize A. Marino$^{5, 9}$(alessio.marino@inaf.it), 
 P.  Marrese$^{5,6}$(paola.marrese@ssdc.asi.it), \\\normalsize R. Molinaro$^2$(roberto.molinaro@inaf.it),  M.I. Moretti$^2$(maria.moretti@inaf.it), 
 \\\normalsize T. Muraveva$^1$(tatiana.muraveva@inaf.it),   V. Ripepi$^2$(vincenzo.ripepi@inaf.it), \\\normalsize  G. Somma$^2$(giulia.desomma@inaf.it), P. Ventura$^5$(paolo.ventura@inaf.it), 
\\\normalsize with the support of the LSST Transient and Variable Stars Collaboration. 
\medskip\\
\footnotesize{$^1$INAF - Osservatorio di Astrofisica e Scienza dello Spazio di Bologna; $^2$INAF - Osservatorio Astronomico} \\\footnotesize{di Capodimonte; $^3$INAF - Istituto di Astrofisica e Planetologia Spaziali, Roma; $^4$Dipartimento di} \\
\footnotesize{Fisica, Universit\`a di Pisa; $^5$INAF - Osservatorio Astronomico di Roma; $^6$Space Science Data Center - ASI;} \\\footnotesize{$^7$Dipartimento di Fisica e Astronomia, Universit\`a di Bologna;}  \\\footnotesize{$^8$Dipartimento di Matematica e Fisica, Universit\'a degli Studi di Roma Tre, Roma;} \\\footnotesize{$^9$Dipartimento di Fisica e Astronomia, Universit\`a La Sapienza, Roma.} }
\date{November 2018}

\begin{document}

\maketitle

\begin{abstract}
This project aims at exploiting the wide-field and  limiting-magnitude capabilities of the LSST to fully characterise the resolved stellar populations 
in/around six  Local Group stellar systems of different morphological type that are located 
from $\sim$30 to $\sim$400 kpc in distance from us.
The adopted stellar tracers will be mainly red giants,  pulsating variable stars of different types, and turn-off  (TO) main sequence (MS) stars.
 In a synergistic and complemental approach, we have selected targets that host red giant branch (RGB) stars  which are within the reach of  {\it Gaia} and not yet (all) saturated  with the LSST. In five of them we will reach  
at least one magnitude below the TO of the oldest  stellar component (t $\ge$ 10 Gyr) with the LSST.
   Specifically, we plan to use pulsating variable stars populating the whole classical
instability strip (namely: RR Lyrae stars, Cepheids of different types, SX Phoenicis and
delta Scuti stars)  and Long Period Variables (LPVs), along with the Color Magnitude Diagram  (CMD)  of the resolved stellar populations in these 6 systems to: 
 i) trace their different stellar generations over the large spatial extension and magnitude  depth allowed by the LSST; ii) measure their distances  
using variable stars of different type/parent stellar population and the Tip of the Red Giant Branch (TRGB); iii) map their 3D structures all the way through to the farther periphery of their halos; 
iv) search for tidal streams which are supposed to connect some of these systems; and v) study their Star Formation Histories (SFHs)  over an unprecedented large fraction of their bodies.  
Our ultimate goal is:  i) to provide a complete picture of these nearby stellar systems all the way through to their periphery, and: ii) to directly link and cross-calibrate the 
{\it Gaia} and LSST projects.
A valuable by-product will be the derivation, for the first time,  of  period-luminosity relations  based on statistically significant samples of delta Scuti and SX Phoenicis stars, in different environments.
\end{abstract}

\section{White Paper Information}
{\bf PIs} and {\bf Contact authors} for this white paper are: Gisella Clementini (gisella.clementini@inaf.it) and Ilaria Musella (ilaria.musella@inaf.it). The full list 
of {\bf Co-authors} is provided above.

\begin{enumerate} 
\item {\bf Science Category:} Structure and Stellar Content of the Milky Way, Exploring the Transient/Variable Universe.
\item {\bf Survey Type Category:} main `Wide-Fast-Deep' survey, minisurvey
\item {\bf Observing Strategy Category:} an integrated program with science that hinges on the combination of pointing and detailed observing strategy. 
\end{enumerate}  

\clearpage

\section{Scientific Motivation}

\begin{footnotesize}
This is a joint proposal of the LSST projects:``The Gaia-LSST
Synergy: from pulsating stars and star formation history to WDs" PI G.
Clementini,  and ``RR Lyrae, Cepheids and Luminous Blue Variables to constrain
theory using LSST observations" PI I. Musella, which optimizes targets and
observing strategy as to meet the goals of both projects. 
In our joint project we propose to observe with the LSST six  Local Group stellar systems of different morphological type,  which are located 
from $\sim$30 to $\sim$400 kpc in distance from us.  In a synergistic and complemental approach, all our selected targets contain red giant branch (RGB) stars  which are within the reach of  {\it Gaia} (e.g. brighter than $G \sim$ 20.7 mag, where $G$ is the {\it Gaia} passband covering  from 330 to 1050 nm), and not yet (all) saturated  with the LSST. {\it Gaia} can reach down to the delta Scuti
stars in Sagittarius, observe RR Lyrae stars and brighter pulsating variables in Sculptor, Carina and, depending on the galaxy actual distance in Antlia 2. In Fornax and Phoenix {\it Gaia} observes 
only bright Cepheids and RGB stars. 
Conversely, the LSST will reach  
at least one magnitude below the TO of the oldest  stellar component (t $\ge$ 10 Gyr) in all these systems but Phoenix. 


The stellar tracers that we plan to use for the characterisation of  our targets include  RGB stars,  pulsating variable stars of different type/parent stellar population that span the whole
classical instability strip (namely: RR Lyrae stars, Cepheids of different types,
SX Phoenicis and delta Scuti stars), LPVs, and  turn-off  (TO) main sequence (MS) stars. 
 In particular, the selected systems include four dwarf spheroidal galaxies (dSphs): Sagittarius, 
 Sculptor, Fornax, Carina,  one dwarf irregular galaxy (dIrr):  Phoenix, and the recently discovered candidate  ``ultra-diffuse"  galaxy 
Antlia~2 (Torrealba et al. 2018).  
Antlia 2 is  located behind the Galactic disc at a galactic latitude of $b \sim 11^{\circ}$  and spans about 1.26 degrees in size, which corresponds to about  2.9 kpc at a distance of 130 kpc estimated for the galaxy in the discovery paper.
 This new stellar system was identified using a combination of astrometry, photometry and variability data (RR Lyrae stars) from Gaia Data Release 2, and its nature confirmed with deep archival DECam imaging, which revealed a conspicuous BHB signal,  in very good agreement with distance the authors estimated from {\it Gaia} RR Lyrae stars.
While similar in extent to the Large Magellanic Cloud, Antlia 2 would be orders of magnitude fainter with $M_{V}$ = $-$8.5 mag, making it by far the lowest surface brightness system known (at 32.3 mag/arcsec$^2$), $\sim$ 100 times more diffuse than the so-called ultra diffuse galaxies. However, controversy exists on  the actual distance of this system because there was an error in the calculation of the distance from the RR Lyrae stars. The {\it Gaia} RR Lyrae stars that triggered the discovery are at about 80 kpc in distance. Therefore, either they do not belong to Antlia 2,  or if they are members the  distance to the galaxy is more on the order of 80-90 kpc, or, as lately suggested by the authors,  the  {\it Gaia} RR Lyrae stars lay in front of the dwarf
along the line of sight and likely represent the near side of an extended cloud of tidal debris spawn by Antlia 2 (V. Belokurov, private communication).
Whatever the correct explanation, the LSST observations of Antlia 2 proposed in our project will definitely shed light  on the nature of this very intriguing new stellar system.

We need  time-series photometry in (at least) three passbands  ($g$, $r$ and $i$, see details below) in order to built multi-band light curves and deep CMDs from which to infer the properties 
of the stellar populations of variable and constant stars  in our targets as well as to derive crucial information on extinction and chemical composition.

All our targets are included in the main ``Wide-Fast-Deep survey" (hereinafter WFD survey) that foresees a sufficient number of visits 
per field per filter to evenly sample the multi-band light curves of the variable stars there contained and allow us an accurate 
determination of their period(s), classification in types and characterization (see below for details). The LMC and SMC are also 
targets of great interest to us. However, the phase coverage currently envisaged  by the South Celestial Pole candidate minisurvey 
(hereinafter SCP; 30 visits per field per filter in ugrizy, in just one visit per night), is insufficient to achieve an accurate 
characterization of the shortest period (e.g. delta Scuti and SX Phoenicis) variable stars in these systems (see section 3.2).  
We note that, the project by Olsen and collaborators  envisages,  among other goals, the observation of RR Lyrae and delta Scuti stars in a number of peripheral fields
of the Magellanic Clouds (MCs).  Synergy between ours and Olsen et al.  project  is possible on the MCs.  

\noindent We note that for all our selected fields/targets the Red Giant Branch (RGB) and  upper Main Sequence (MS) stars are within the reach of  Gaia and not yet (all) saturated with the LSST\footnote{RGB and MS stars up to about 1.6 mag brighter than the horizontal branch of Sagittarius are still below the saturation limit of the LSST, $r \sim$ 16 mag} thus  allowing to directly link and cross-calibrate the two surveys/missions.

The proposed project will allow us to: 
\begin{itemize}
\item
intercalibrate the LSST and Gaia datasets using primarily variable stars, but also constant RGB/MS stars.  The {\it Gaia}-LSST calibration on variable stars will be a crucial step 
for using variables discovered by the LSST as distance indicators and to probe stellar pulsation models such as, for instance, those in Marconi et al. 2015; 
\item
optimally translate into the LSST passbands  different theoretical and empirical 
diagnostic tools that, as members of Coordination Unit 7 (variability) in the Gaia Data Processing and Analysis Consortium (DPAC), we specifically developed  and fine tuned for the processing and characterization of the aforementioned variability types  (see e.g. Clementini et al. 2016, 2018; Ripepi et al. 2018); 
\item
test  the quality assurance of  products for the pulsating stars (light curves, pulsation parameters etc.); 
\item
compare the LSST magnitude limits and performance with respect to {\it Gaia} in regions of high crowding/absorption;
\item
define period-luminosity relations  based on statistically significant samples of delta Scuti and SX Phoenicis stars, in different environments (see, e.g. Poretti et al. 2008, for a first 
example); 
\item
test and validated our nonlinear convective pulsation models for several classes of pulsating stars, including in particular, delta Scuti, SX Phoenicis and LPVs  (see e.g. McNamara et al. 2007; Trabucchi et al. 2017 and references therein);

\item
 trace the different stellar generations of the selected systems over the large spatial extension and depth that the LSST allows us to achieve; 
\item
 measure their distances  using variable stars of different type/parent stellar population and the Tip of the Red Giant Branch (TRGB)
\item
 map their 3D structures all the way through to the farther periphery of their halos; 
\item
search for tidal streams which are supposed to connect some of the selected  systems; 
\item
test depth and completeness of the LSST observations in the selected  systems/fields and study the Star Formation Histories (SFHs)  over unprecedented large fractions of their bodies.  
\end{itemize}

This  project will provide a new, complete picture of these nearby stellar systems all the way through to their periphery,  by benefitting from the combination of  
{\it Gaia} and LSST observations (including parallaxes and proper motions), with the variable stars and  SFH recovery approaches (see e.g. Clementini et. al. 2012, Cignoni et al. 2018).

We run the LSST  metrics to test the recovery performance for the various types of variable stars we plan to observe in our targets, (see details in the Technical Description). The recovery performance was specifically computed in fields centred on our six targets, results are summarised in Table~\ref{numobs}. 

Initially,  we had planned  to observe three additional fields around M54 to  better sample the significantly large extension of the Sagittarius dSph.  However,  the number of observations foreseen within the first two years by the baseline2018 for regions outside known stellar systems remains  uncomfortably small even if observations were doubled. Therefore, we only ask for doubling the number of observations within the first two years for the fields centred on our targets (see  Technical Description) as we want to be able to accurately recover  characteristic  parameters for the variable stars in these central regions. For the external regions, in the first two years, we will only detect  candidate variables. To this purpose, without any modification of the cadence, even in present of only a few epoch-data, we will detect candidate variables using semi-supervised classification approaches as described in detail in Rimoldini et al. (2018) for the classification of variable stars observed by Gaia with less than 12 epochs. A similar approach can be applied to the external fields of all stellar systems selected in this project, particularly if the first two year analysis will allow us to detect stellar over-densities that are likely the signature of  tidal radii  and streams extending out of the central regions of these systems.


As an additional  significant added value, the observation in our project  of different types of pulsating variables tracing the different stellar populations in the host systems will allow us to 
directly cross-calibrate  Population I (Pop I; e.g. Classical Cepheids, delta Scuti stars) and Population II (Pop II; e.g. RR Lyrae and SX Phoenicis stars) standard candles and distance indicators (e.g. the tip of red giant branch) at same time and in a fully homogeneous way.  

\end{footnotesize}

\vspace{.6in}

\section{Technical Description}

\subsection{High-level description}
\begin{footnotesize}

We propose a combination of the Wide-Fast-Deep survey and a minisurvey (or alternatively a modified WFD with a rolling cadence that doubles the number of visits in the first two years), 
in order to study the stellar variability of a selected sample of stellar systems of different morphological type.\\
The cadence of the WFD survey simulated in the recent baseline2018a.db foresees two visits per field (either in the same or in different filters) in the same night. These pairs of visits are repeated every three to four nights throughout the period the field is visible in each year, for a total of 825 visits over the 6 filters: $ugrizy$. \\

\end{footnotesize}

\begin{table}[ht]
\caption{Information on our selected stellar systems} \label{numobs}
    \centering
    {\small
    \begin{tabular}{|l|l|l|l|l|l|l|l|}
        \toprule
        Target & RA(J2000)  &DEC(J2000)& $(m-M)_0$ & E(B-V) & Ng & Nr & Ni \\
        \midrule
M54/Sagittarius &18$^h$55$^m$03.3$^s$ & $-$30$^{\circ}$28$^m$43.0$^s$& 17.13&0.15&13&33&37\\
Sculptor   & 01$^h$00$^m$09.35$^s$ & $-$33$^{\circ}$42$^m$32.5$^s$&19.57&0.016&26&44&47\\ 
Carina     & 06$^h$41$^m$36.7$^s$    & $-$50$^{\circ}$57$^m$58.0$^s$&20.08&0.05&14&33&31\\
Fornax     & 02$^h$39$^m$59.33$^s$ & $-$34$^{\circ}$26$^m$57.1$^s$&20.70&0.02&28&50&56\\
Phoenix  & 01$^h$51$^m$06.3$^s$ &  $-$44$^{\circ}$26$^m$41$^s$&23.10&0.014&14&23&24\\
Antlia 2  & 9$^h$3$5^m$32.64$^s$  &$-$36$^{\circ}$46$^m$02.28$^s$&20.56 &0.19 &26&57&58\\
        \bottomrule
    
    \end{tabular}
    }
\end{table}

\begin{figure}
    \centering
    \includegraphics[width=18 cm, trim=100 100 100 100 mm]{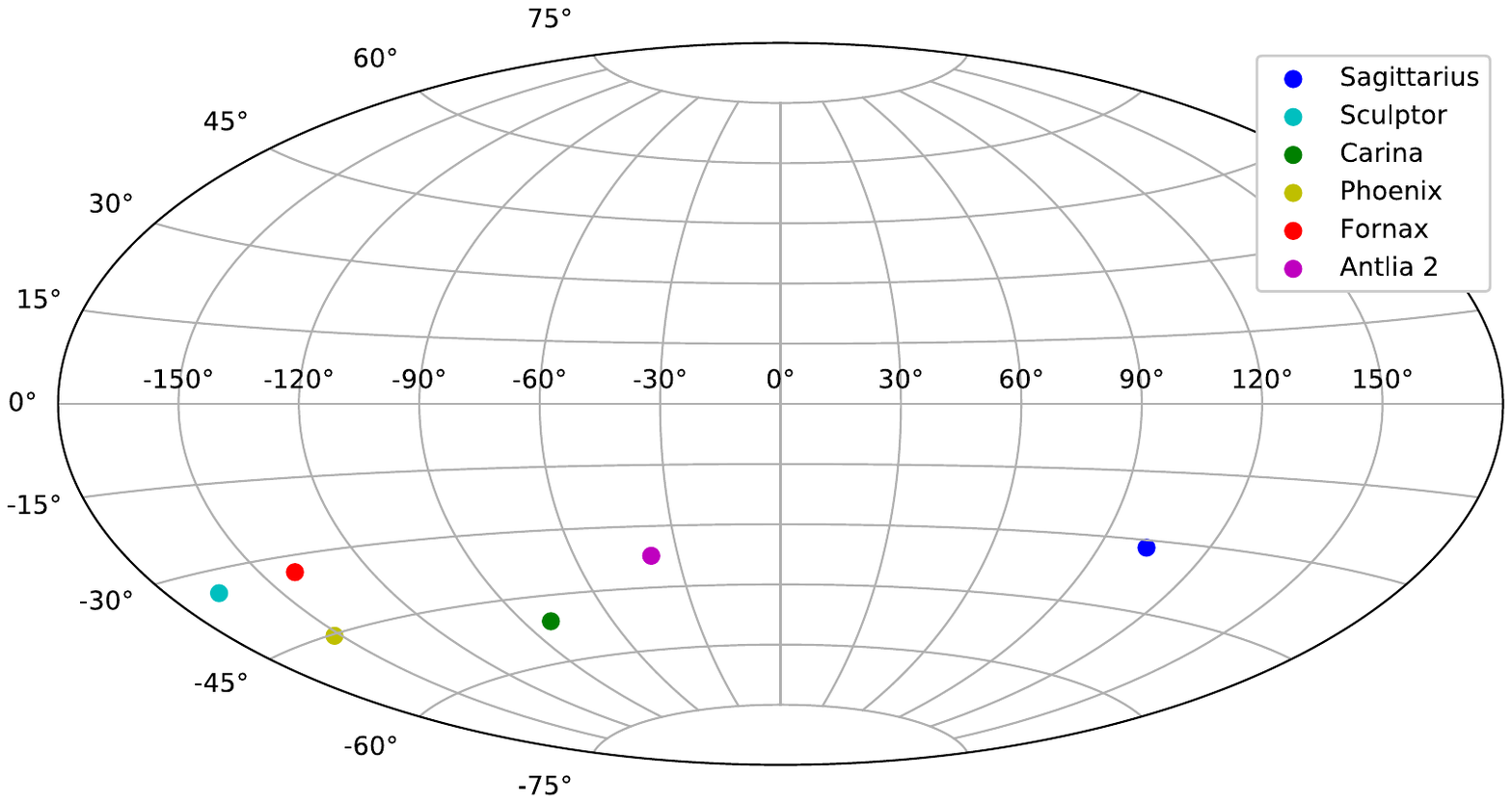}

    \caption{Distribution on the sky of the six stellar systems/LSST fields targeted in the proposal.}\label{posgal}
\end{figure}

\begin{figure}
    \centering
    \includegraphics[width=15 cm]{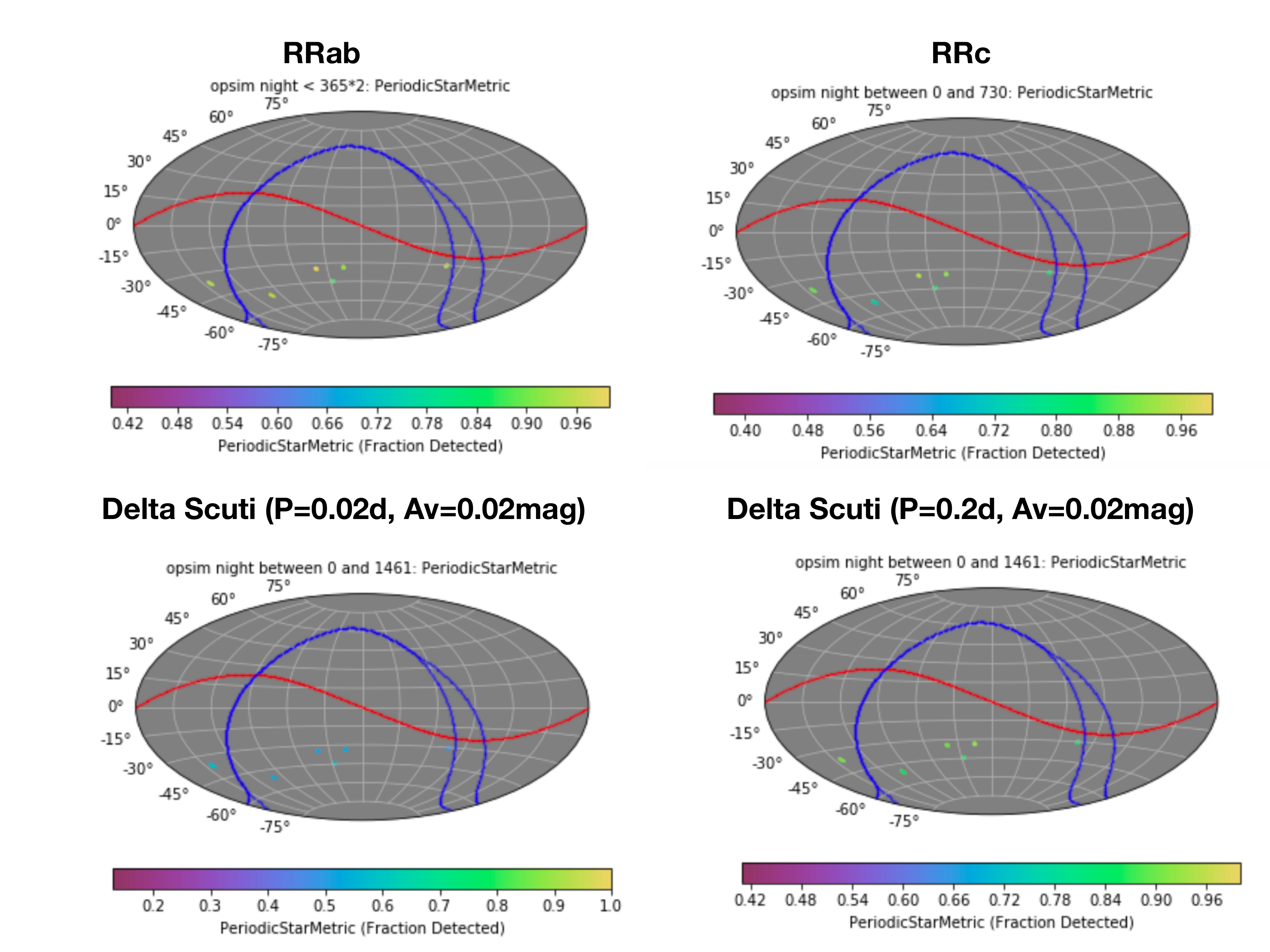}
    \caption{Recovered fraction of different types of RR Lyrae and delta Scuti stars detected in the first two years. See text for details.}
    \label{recovery}
\end{figure}

\begin{footnotesize}

In Table~\ref{numobs} we list  the selected stellar systems, their coordinates, distance modulus and color excess. Observation 
of all our targets is included in the WFD survey and according to the baseline2018 they are reached by a different number of visits. In particular, in the last three columns of the Table \ref{numobs} we have reported the number of visits in the $g$, $r$ and $i$ bands (N$g$, N$r$ and N$i$), planned in the first two years, for fields that include the center of these six systems. These fields are showed in Fig. \ref{posgal}. To get this information about fields and visits we developed a very simple notebook 
($https://bit.ly/2QuNieS$).
Another notebook 
($ https://bit.ly/2Quvcd1$) 
 uses an existing metric (PeriodicStarMetric.py) to get the fraction of recovered  variable stars based on the period, amplitude and mean magnitude (in $ugriyz$) of the different types of variable stars.  
To properly take into account the period, amplitude and magnitude ranges spanned by RR Lyrae and delta Scuti stars, we built a grid of possible periods, amplitudes and mean magnitudes on which to apply the metric PeriodicStarMetric.py. This grid was built for the LMC RR Lyrae stars using the period-luminosity ($PL$) relations in the ($V$) and $I$ bands and the Period-amplitude relation in the $I$ band obtained by Soszy{\'n}ski et al.(2009) and for the LMC delta Scuti and SX Phoenicis stars using the $PL$ relation by McNamara et al. (2007). delta Scuti  stars do not obey to a period/amplitude relation and their amplitude spans from 0.02 to 0.2 mag (see e.g. Poretti et al., 2008, ApJ, 685, 947). Their magnitudes have been de-reddened, transformed to the $ugrizy$ bands (Jordi et al., 2005) and placed at the distance of the selected stellar systems. 
We find that the fraction of recovered  variable stars strongly depends on the galaxy distance, but also on the magnitude and amplitude of the variables (see Fig. \ref{recovery}). Specifically,  for the first two years of observation, the degree of recovery for c-type RR Lyrae  stars is 100\% for M54, Sculptor and Carina, $\sim$ 90\% for Fornax and Antlia2 and $\sim$ 50\% for Phoenix. High amplitude RRab stars  are recovered at 100\% in all systems except Phoenix (with 67\%), but for the low amplitude ones the recovery level goes down to 90\% for Sculptor and Carina, 80\% for Fornax and Antlia and to 56\% for Phoenix. For delta Scuti stars the recovering level is much lower because they are significantly fainter stars. In particular, going from low to high amplitude delta Scuti,  for example in M54 we have a recovery ranging from 50\% to 60\%. For  low amplitude delta Scuti stars, in the other systems, we have a maximum recovery level of 10-20\% and even much lower in Phoenix, whereas for the high amplitude Delta Scuti stars, the level increases to 30\%-50\%.

To properly take into account the previous results we have to note that the adopted metric represents only a first approximation of the light curve of our variables and in particular it does not consider the multi-periodic nature of the delta Scuti variables.
Many proponents of this project are members of Coordination Unit 7 (Variability) in the {\it Gaia} Data Processing and Analysis Consortium (DPAC) and in particular G. Clementini is the responsible for RR Lyrae and Cepheids pipeline. Based on the expertise gained from the processing of variable stars observed by  {\it Gaia}, we judge that to measure accurate periods, amplitudes and mean magnitudes, we need at least about 30  phase points for RR Lyrae stars  and 60 epochs for delta Scuti stars. For this reason, we propose to double the scheduled cadence  on our targets in the first two years at least in the $g$, $r$ and $i$ bands. We adopt these filters because the detection of these types of variables in the near-infrared $z$ and $y$ passbands  where their amplitude is much smaller is very difficult or even impossible, 
regardless the much higher number of visits expected in these bands.

The two possible strategies are either to double the number of visits on the selected systems in the first 2 years by then reducing the number in  subsequent years, or to add the necessary visits on these systems in a minisurvey framework. 
Summarizing, the proposed strategies allow us to optimize the coverage of short period (P$<$1 d) 
variables such as RR Lyrae, delta Scuti and SX Phoenicis stars in the selected stellar systems. However,  we also need the observations provided in subsequent years in the main WDF survey to study secondary modulations on a timescale longer than main oscillations (e.g. the Blazko effect in the RR Lyrae stars, Blazhko 1907) and also to detect and characterise longer period variables.

\end{footnotesize}

\vspace{.3in}

\subsection{Footprint -- pointings, regions and/or constraints}
\begin{footnotesize}

Targets/fields of our project are included in the WFD survey and  listed in Table \ref{numobs}.

The LMC and SMC would also be very interesting targets for our project,  in particular, to study the delta Scuti/SX Phoenicis populations, which are below the limiting magnitude of current surveys 
(e.g. OGLE ). However, the fields on and around the MCs in baseline2018 are  included in the South Pole Celestial minisurvey with a coverage much lower respect to WFD survey. 
Furthermore, the project "Mapping the Periphery and Variability of the Magellanic Clouds" by Olsen and collaborators,  proposes of doubling the number of visits in the first two years to increase the recovery of short-term variables in the MCs. Synergy between ours and Olsen et al. projects is definitely possible.

\end{footnotesize}

\subsection{Image quality}
\begin{footnotesize}{We need a seeing $<$ 1.0 arcsec for the observation of crowded fields such as the centre of M54 or the Phoenix dIrr galaxy, while seeing $<$ 1.5 arcsec is adequate for our other targets . 
}
\end{footnotesize}

\subsection{Individual image depth and/or sky brightness}
\begin{footnotesize}{Our strategy has been based on the image depth and/or sky brightness provided by baseline2018a.
}
\end{footnotesize}

\subsection{Co-added image depth and/or total number of visits}
\begin{footnotesize}{The two different proposed strategies are either to double the number of visits on the selected systems in the first 2 years by then reducing the number in  subsequent years, or to add the necessary visits on these systems in a minisurvey framework.}
\end{footnotesize}

\subsection{Number of visits within a night}
\begin{footnotesize}{Two visits per night in the same filter.
}
\end{footnotesize}

\subsection{Distribution of visits over time}
\begin{footnotesize}{ 
The two possible strategies are to increase (at least doubling) the number of observation on the selected systems (Table \ref{numobs}) in the first 2 years reducing the number in the subsequent ones, or to add the necessary observations on these systems in a minisurvey framework. 
These proposed strategies allow to optimize the coverage of short period (P$<$1 d) 
variables such as RR Lyrae, delta Scuti and SX Phoenicis stars in the selected stellar systems, but we also need the observations provided in the subsequent years in the main WDF survey to study secondary modulations on a timescale longer than main oscillations (Blazko effect in the RR Lyrae light curves for example) and also to detect and study longer period variables.

}
\end{footnotesize}

\subsection{Filter choice}
\begin{footnotesize}
We propose to double $g$, $r$ and $i$ observations in the first two years on the selected stellar systems (Table \ref{numobs})
\end{footnotesize}

\subsection{Exposure constraints}
\begin{footnotesize}
We assume the  15x2-second visits with double snaps as adopted in the baseline2018a. We prefer 15-second exposure instead of 30 in order to sample rapid timescale variability and to limit saturation.
\end{footnotesize}

\subsection{Other constraints}
\begin{footnotesize}
{\it 
}
\end{footnotesize}

\subsection{Estimated time requirement}
\begin{footnotesize}
To double the number of $g$, $r$ and $i$ visits  (with respect to what envisaged by the WFD survey) in the first  two
years, if we adopt the minisurvey strategy, we need additional 614
visits of all the selected fields. Given that a 10\% of the survey time
is reserved for Deep Drilling and minisurvey ($\sim$260,000 visits),
this means that our minisurvey of dwarfs uses only 614/260000=0.25\% of
the time available for DD and minisurvey
\end{footnotesize}

\vspace{.3in}

\begin{table}[ht]
    \centering
    \begin{tabular}{l|l|l|l}
        \toprule
        Properties & Importance \hspace{.3in} \\
        \midrule
        Image quality &  1   \\
        Sky brightness &  2 \\
        Individual image depth & 1   \\
        Co-added image depth &  3 \\
        Number of exposures in a visit  & 1  \\
        Number of visits (in a night)  & 1  \\ 
        Total number of visits & 2  \\
        Time between visits (in a night) & 2 \\
        Time between visits (between nights)  & 2  \\
        Long-term gaps between visits & 3 \\
        Other (please add other constraints as needed) & \\
        \bottomrule
    \end{tabular}
    \caption{{\bf Constraint Rankings:} Summary of the relative importance of various survey strategy constraints. Please rank the importance of each of these considerations, from 1=very important, 2=somewhat important, 3=not important. If a given constraint depends on other parameters in the table, but these other parameters are not important in themselves, please only mark the final constraint as important. For example, individual image depth depends on image quality, sky brightness, and number of exposures in a visit; if your science depends on the individual image depth but not directly on the other parameters, individual image depth would be `1' and the other parameters could be marked as `3', giving us the most flexibility when determining the composition of a visit, for example.}
        \label{tab:obs_constraints}
\end{table}

\subsection{Technical trades}
\begin{footnotesize}
{\it To aid in attempts to combine this proposed survey modification with others, please address the following questions:}
\begin{enumerate}
    \item {\it What is the effect of a trade-off between your requested survey footprint (area) and requested co-added depth or number of visits?}\\
    
    We are interested in the co-added images to built a color-magnitude diagram as deeper as possible to recover the star formation history of the studied stellar system 
    
    \item {\it If not requesting a specific timing of visits, what is the effect of a trade-off between the uniformity of observations and the frequency of observations in time? e.g. a `rolling cadence' increases the frequency of visits during a short time period at the cost of fewer visits the rest of the time, making the overall sampling less uniform.}\\
    
    A change of the rolling cadence is one of the two strategies proposed in this project.
    
    \item {\it What is the effect of a trade-off on the exposure time and number of visits (e.g. increasing the individual image depth but decreasing the overall number of visits)?}\\
    
    We prefer to increase the number of visits and not increase the exposure time to avoid the saturation of the brightest objects.
    
    \item {\it What is the effect of a trade-off between uniformity in number of visits and co-added depth? Is there any benefit to real-time exposure time optimization to obtain nearly constant single-visit limiting depth?}\\
    
    We have no constraints on this point.
    
    \item {\it Are there any other potential trade-off to consider when attempting to balance this proposal with others which may have similar but slightly different requests?}
    
    We have specified our necessity in the Technical Description section.
    
\end{enumerate}
\end{footnotesize}

\section{Performance Evaluation}
\begin{footnotesize}
The Metrics used  for this project are already present in  /home/docmaf/maf local/
sims\_maf\_contrib/mafContrib/. In  particular periodicStarMetric.py was
used in the two notebooks that we have uploaded on GitHub. For this
work also other
already written  notebooks  were checked such as
slowTransientsVariables/Variability Metrics.ipynb,
variabilityDepth/variabilityDepth.ipynb and
Transients/TransientAsciiMetric.ipynb

\end{footnotesize}

\vspace{.6in}

\section{Special Data Processing}
\begin{footnotesize}
Detection of variable stars through image subtraction and PSF photometry are  our optimal data processing procedures. Image subtraction and PSF photometry 
will be performed as part of Prompt processing. Therefore this program does not require any special processing beyond what DM already plans to provide.

\end{footnotesize}

\section{Acknowledgments}
\begin{footnotesize}
This work developed partly within the TVS Science Collaboration and the author acknowledge the support of TVS in the preparation of this paper. MDC thanks the organizers of LSST cadence Hackathon at Flatiron Institute and LSST Corporation and the Simons Foundation for their support.
\end{footnotesize}

\section{References}
\begin{footnotesize}
Blazko, S. 1907, AN, 175, 325\\
Cignoni, M., Sacchi, E., Aloisi, A., et al.\ 2018, ApJ, 856, 62\\
Clementini, G., Cignoni, M.. Contreras Ramos, R., et al. 2012,  ApJ, 756, 108\\ 
Clementini, G., Ripepi, V., Leccia, S., et al.\ 2016, A\&A, 595, A133\\
Clementini, G., Ripepi, V., Molinaro, R., et al.\ 2018, arXiv:1805.02079\\ 
Jordi, K., Grebel, E.~K., \& Ammon, K.\ 2005, Astronomische Nachrichten, 326, 657\\ 
Marconi, M., Coppola, G., Bono, G., et al.\ 2015, ApJ, 808, 50\\ 
McNamara, Clementini \& Marconi\ 2007, AJ, 133, 2752\\ 
Poretti, E., Clementini, G., Held, E.~V., et al.\ 2008, ApJ, 685, 947 \\
Ripepi, V., Molinaro, R., Musella, I., et al.\ 2018, arXiv:1810.10486 \\ 
Soszy{\'n}ski, I., Udalski, A., Szyma{\'n}ski, M.~K., et al.\ 2009, ACA, 59, 1\\
Rimoldini, L.,  Holl, B.,  Audard, M., et al.\ 2018, arXiv:1811.03919\\
Torrealba, G., Belokuro, V., Koposov, S.E., et al.\ 2018, arXiv:1811.04082\\
Trabucchi, M., Wood, P.~R., Montalb{\'a}n, J., et al.\ 2017, ApJ, 847, 139 \\
\end{footnotesize}

\end{document}